\documentclass[fleqn,twoside]{article}
\usepackage{espcrc2}
\usepackage{amssymb}

\usepackage{graphicx}
\usepackage[figuresright]{rotating}

\newcommand{\AmS}{{\protect\the\textfont2
  A\kern-.1667em\lower.5ex\hbox{M}\kern-.125emS}}

\title{Non-commutative Solitons in Finite Quantum Mechanics}

\author{E. G. Floratos\address[INP]{INP NRCPS ``Demokritos''\\  
        15310 Aghia Paraskevi, Athens, Greece\\ and\\
Nuclear and Particle Physics Physics Section, University of Athens\\
15771 Athens, Greece}
        \thanks{E-mail: Manolis.Floratos@inp.demokritos.gr}
and
        S. Nicolis\address{CNRS--Laboratoire de Math\'ematiques et
        Physique Th\'eorique (UMR 6083) \\
        Universit\'e de Tours, Parc Grandmont 37200 Tours, France}
        \thanks{Speaker. E-mail: Stam.Nicolis@phys.univ-tours.fr}
}
       
\begin{document}

\begin{abstract}
We construct the unitary evolution operators that realize the
quantization of linear maps of $SL(2,{\mathbb{R}})$ over phase spaces
of arbitrary integer discretization $N$ and show the non-trivial
dependence on the arithmetic nature of $N$. We discuss the
corresponding uncertainty principle and construct the corresponding
coherent states, that may be interpreted as non-commutative solitons.
\vspace{1pc}
\end{abstract}

\maketitle

\section{The Heisenberg--Weyl Group and its Consistent Discretization}
In gauge theories in the continuum the fields take values in the
corresponding  Lie algebra, 
$$
A_\mu(x)=A_\mu^{a}(x)T_a
$$
The theory is invariant under Lorentz transformations and under gauge
trasnformations. 

In lattice gauge theories the fields take values in the {\em group}
corresponding to the Lie algebra 
$$
U_\mu(x)=\exp({\mathrm{i}}A_\mu(x))
$$
The continuous Lorentz symmetry is broken to a discrete subgroup--but
the internal (gauge) symmetry is preserved. Note that it is this
feature that renders the study of lattice supersymmetry so
non-trivial, since supersymmetry mixes Lorentz symmetry and internal
symmetry in a non-trivial way. 

In quantum mechanics we consider functions that take values in phase
space. These functions are valued in the Heisenberg algebra
\begin{equation}
\label{heisenberg}
\left[q,p\right]={\mathrm{i}}\hbar I
\end{equation}
This algebra has only infinite dimensional representations, so
discretization of the phase space does not preserve this symmetry. It
is known since Weyl, however, that the exponentials of these
generators, 
\begin{equation}
\label{weyl}
\begin{array}{lr}
P=\exp({\mathrm{i}}p) & Q=\exp({\mathrm{i}}q)
\end{array}
\end{equation}
do admit finite--dimensional representations, that respect the 
group law
\begin{equation}
\label{grouplaw}
QP=\underbrace{e^{{\mathrm{i}}2\pi/N}}_{\omega_N}PQ
\end{equation}
for an $N-$dimensional representation. 

The generators of the Heisenberg--Weyl group are the displacement
operators
\begin{equation}
\label{groupgen}
J_{r,s}=\omega_N^{r\cdot s/2}P^rQ^s
\end{equation}
where $(r,s)$ denotes the point on the classical discretized phase
space. 

The strategy will now be to write the evolution operator that realizes
a consistent quantization of the classical motion. This will turn out
to involve the arithmetic nature of the discretization in a
non-trivial way. One thus hopes to extract non-trivial information, in
the same way the continuum limit of a lattice field theory is
non-trivial, if a second order phase transition occurs. 

In what follows we will highlight, through explicit examples, the
non-trivial $N-$dependence of the evolution operator and indicate how one 
may also compute correlation functions. 

We will further construct a subset of states that are permuted among
themselves by certain evolution operators and whose dispersion in
position and momenta remain constant in time--they may thus be 
called solitons. They depend non-trivially on the discretization,
$N$. Recent work seems to hint towards a close relation with the
solitons of non-commutative scalar field theory~\cite{chinese}.

\section{The Evolution Operator on a Discrete Phase Space}
We work on toroidal phase spaces, where a convenient parametrization of the
Hilbert space of functions is given by 
\begin{equation}
\label{torusbas}
f(z|\tau)=\sum_{n\in{\mathrm{N}}}c_ne^{{\mathrm{i}}\pi n^2\tau+2\pi{\mathrm{i}}nz}
\end{equation}
On these functions we define the action of the following operators
\begin{equation}
\label{torusheisweyl}
\begin{array}{l}
S_bf(z)= f(z+b) \\
 T_af(z)=e^{{\mathrm{i}}\pi a^2\tau+2\pi{\mathrm{i}}az}f(z+a\tau) 
\end{array}
\end{equation}
and we may easily verify that they realize an infinite dimensional
representation of the Heisenberg--Weyl group. Taking periodic
coefficients, $c_n=c_{n+N}$ leads to an $N-$dimensional
representation. 

Using the formalism set forth in refs.~\cite{dikamas0,dikamas1}, we
obtain the operator, that realizes the quantization of the classical
motion, i.e. the quantum evolution operator, in the form 
\begin{equation}
\label{evolution}
U({\sf A})_{k,l}=c_N({\sf A})\omega_N^{A_1k^2+A_2kl+A_3l^2}
\end{equation}
where the indices $k$ and $l$ take the values $0,1,\ldots,N-1$.

The coefficients $A_{1,2,3}$ as well as the prefactor $c_N({\sf A})$
are fixed by demanding that 
\begin{itemize}
\item $$U({\sf A}\cdot {\sf B})=U({\sf A})U({\sf B})$$
for all matrices ${\sf A}$ and ${\sf B}$ of $SL(2,{\mathrm{Z}}_N)$
i.e. that they be a group representation; and
\item $$J_{(r,s){\sf A}}U({\sf A})=U({\sf A})J_{r,s}$$
i.e. that they enjoy the {\em metaplectic} property. 
(it is interesting to note that the second implies the first)
\end{itemize}
We find $A_1=a/(2c)$, $A_2=-1/c$, $A_3=d/(2c)$ and $c_N({\sf
A})=\sigma_N(1)\sigma_N(2-a-d)/N$, where
$\sigma_N(a)=(1/\sqrt{N})\sum_{r=0}^{N-1}\omega_N^{ar^2}$ is 
the quadratic Gauss sum\cite{lang}.
All operations are taken mod $N$. 

\section{Solitons on ${\mathbb{T}}^2/{\mathbb{Z}}_N$}
Let us now pass to issues of dynamics. The evolution operator acts on
states and one is particularly interested in studying their dispersion
with time. An interesting classical motion is a {\em dilatation},
where $a=1/d$ and $b=c=0$ (mod $N$). Using the results of
\cite{dikamas0,dikamas1} it is possible to show that the corresponding
quantum operator is proportional to $\delta_{k,k'a}$, i.e. a
permutation operator. We thus find that this classical motion leads to 
states that are non-dispersive in both position and momentum--they are
solitons. All classical motions, whose matrix ${\sf A}\in
SL(2,{\mathbb{Z}}_N)$ is diagonalizable mod $N$ describe such
states. An example, for $N=4l+1$, is the classical
rotation\cite{floleon}, since
$$
\left(\begin{array}{cc} a & b \\ -b & a\end{array}\right)=
{\sf V}\left(\begin{array}{cc} a-tb & 0 \\ 0 &
a+tb\end{array}\right){\sf V}^{-1}
$$
with
$$
{\sf V}=\frac{1}{1+t}\left(\begin{array}{cc}  1 & 1 \\  -t &
t\end{array}\right)
$$
and
$$
t\equiv q^l\,{\mathrm{mod}}\,N
$$

where $q$ is the generator of ${\mathbb{Z}}/{\mathbb{Z}_N}$.

\section{Correlation Functions}
Using Gauss sum technology it is possible to evaluate correlation
functions. Indeed, any operator may be written as a linaer combination
of the Heisenberg-Weyl group generators $J_{r,s}$ so, for integer
values of the inverse temperature on may readily evaluate 
$$
\frac{ {\mathrm{Tr}} [J_{r1,s1}\cdots J_{r_k,s_k}U({\sf A})^\beta]}{ {\mathrm{Tr}} [U({\sf A})^\beta] }
$$
These computations are similar to the so-called {\em number theoretic}
quantum spin chains studied by Knauf and collaborators\cite{knauf}.

\section{Conclusions and Outlook}
Working directly in phase space, instead of configuration space leads
to alternative quantization schemes, that go back to Weyl and Wigner. 
One advantage is the natural existence of finite dimensional
representations, that respect the group structure. While the continuum
limit is a non-trivial one, as it involves resumming oscillating
terms (cf. however ref.\cite{vhf})
the formalism holds promise of naturally including features,
hard to describe by other approaches. 

An appropriate example is in the
recent proposal for supersymmetry on a spatial lattice\cite{kaplan},
where the lattice is indeed a phase space of a quantum mechanical
system.

\end{document}